\documentclass[11pt,a4paper]{article}
\pdfoutput=1
\usepackage{jheppub}
\usepackage{url}
\usepackage{tikz}
\usepackage{subcaption}
\DeclareMathOperator{\Tr}{Tr}

\newcommand{\hf}{\frac{1}{2}}

\newcommand{\del}{\partial}

\newcommand{\bra}{\langle}
\newcommand{\ket}{\rangle}

\newcommand{\bt}{\beta}
\newcommand{\ga}{\gamma}
\newcommand{\Ga}{\Gamma}
\newcommand{\al}{\alpha}

\newcommand{\rt}[1]{\sqrt{#1}}
\newcommand{\cO}{\mathcal{O}}

\begin{document}

\title{Capacity of entanglement in random pure state}

\author{Kazumi Okuyama}

\affiliation{Department of Physics, Shinshu University,\\
3-1-1 Asahi, Matsumoto 390-8621, Japan}

\emailAdd{kazumi@azusa.shinshu-u.ac.jp}

\abstract{
We compute the capacity of entanglement 
in the bipartite random pure state model
using the replica method.
We find the exact expression of the capacity of entanglement
which is valid for a finite dimension of the Hilbert space.
We argue that in the gravitational path integral, the capacity of entanglement
receives contributions only from the sub-leading saddle points corresponding to
the partially connected geometries.
}

\maketitle

\section{Introduction}
In recent papers \cite{Penington:2019kki,Almheiri:2019qdq},
the Page curve of the Hawking radiation \cite{Page:1993wv}
is reproduced from the replica computation of the 
entanglement entropy (see also \cite{Almheiri:2020cfm}
for a review).
As argued by Page in \cite{Page:1993df}, 
the entropy computation of the Hawking radiation is nicely modeled
by the random pure state $|\Psi\ket$ in a bipartite Hilbert space 
\begin{equation}
\begin{aligned}
 \mathcal{H}=\mathcal{H}_A\otimes \mathcal{H}_B.
\end{aligned} 
\label{eq:HA-HB}
\end{equation}
Here the subsystems $A$ and $B$ correspond to
the Hawing radiation and the black hole, respectively.
From the reduced density matrix $\rho_A$ of the subsystem $A$
\begin{equation}
\begin{aligned}
 \rho_A=\Tr_B|\Psi\ket\bra\Psi|,
\end{aligned} 
\label{eq:rhoA}
\end{equation}
we can compute the entanglement entropy $S_A$
\begin{equation}
\begin{aligned}
 S_A=-\bra \Tr\rho_A\log\rho_A\ket,
\end{aligned} 
\end{equation}
where the bracket $\bra\cdots\ket$ denotes the ensemble average over the random
pure state $|\Psi\ket$.
The exact form of $S_A$ \eqref{eq:SA-exact} is obtained 
in \cite{Page:1993df} as a function of the dimensions $d_A,d_B$
of the Hilbert spaces \eqref{eq:HA-HB}
\begin{equation}
\begin{aligned}
 d_A=\dim\mathcal{H}_A,\quad
d_B=\dim\mathcal{H}_B.
\end{aligned} 
\label{eq:dA-dB}
\end{equation}
As discussed in \cite{Page:1993wv}, $S_A$ in the random pure state
model exhibits a similar behavior as the Page curve for the Hawking radiation
from an evaporating black hole.

In a recent paper \cite{Kawabata:2021hac}, 
it is argued that the capacity of entanglement $C_A$ introduced in \cite{qi}
is a useful quantity to diagnose the phase transition
around the Page time.
$C_A$ is defined by
\begin{equation}
\begin{aligned}
 C_A=\Tr(\rho_A K^2)-(\Tr\rho_A K)^2,
\end{aligned} 
\end{equation} 
where $K=-\log\rho_A$ is the modular Hamiltonian.
In other words, $C_A$ measures the fluctuation of the modular Hamiltonian.
See e.g. \cite{Nakaguchi:2016zqi,Nakagawa:2017wis,deBoer:2018mzv} for the study of
$C_A$ in various models.

In this paper, we compute the capacity of entanglement in the random pure state model
using the replica method.
From the ensemble average
of $\Tr\rho_A^n$ over the random pure state, 
we can compute the entanglement entropy $S_A$
and the capacity of entanglement $C_A$ as
the derivative with respect to the replica number $n$ at $n=1$
\begin{equation}
\begin{aligned}
 S_A&=-\del_n\log\bra\Tr\rho_A^n\ket\Big|_{n=1}=
-\del_n\bra\Tr\rho_A^n\ket\Big|_{n=1},\\
C_A&=\del_n^2\log\bra\Tr\rho_A^n\ket\Big|_{n=1}=
\del_n^2\bra\Tr\rho_A^n\ket\Big|_{n=1}-(S_A)^2.
\end{aligned} 
\label{eq:def-SC}
\end{equation}
We find the exact form of $C_A$ as a function of the dimensions $d_A,d_B$ of the Hilbert
spaces $\mathcal{H}_{A},\mathcal{H}_{B}$.
The exact expression of $C_A$ in \eqref{eq:C-exact}
is the main result of this paper.

As discussed in \cite{Penington:2019kki,Almheiri:2019qdq},
fully disconnected or fully connected geometries
dominate in the replica computation of the entropy $S_A$, and their 
contributions exchange dominance around the Page time.
In the case of the capacity $C_A$, it turns out that
the leading saddle point from the fully disconnected or fully connected geometries
does not contribute to $C_A$, and it receives non-zero contributions only from the 
sub-leading saddle points corresponding to the partially connected geometries.  
This is consistent with the
result of \cite{deBoer:2018mzv} that $C_A$ is a measure of the partial entanglement. 

This paper is organized as follows.
In section \ref{sec:random}, we review the random pure state model
and the known exact result of $\bra\Tr\rho_A^n\ket$.
We find a new formula of $\bra\Tr\rho_A^n\ket$ in terms of the 
Narayana number \eqref{eq:tr-narayana}, which is useful 
for the replica computation of the entanglement entropy $S_A$
and the capacity of entanglement $C_A$.
In section \ref{sec:planar}, we compute $S_A$ and $C_A$ in 
the planar limit using the replica method.
Our computation shows that the leading saddle point does not contributes to
$C_A$ and it receives contributions only from sub-leading saddle points. 
In section \ref{sec:exact}, we compute the exact $S_A$ and $C_A$ 
using the replica method. 
Finally we conclude in section \ref{sec:conclusion}.

\section{Random pure state model}\label{sec:random}
In this section, let us briefly review the random pure state model.
We consider a pure state $|\Psi\ket$ in the bipartite Hilbert space 
$\mathcal{H}=\mathcal{H}_A\otimes \mathcal{H}_B$.
This models the black hole evaporation where 
$A$ corresponds to the Hawking radiation while $B$ corresponds to the black hole.
In the model discussed in \cite{Penington:2019kki},
$A$ and $B$ correspond to the end of the world brane and the bulk 
JT gravity, respectively. 
We can expand the state $|\Psi\ket$ in terms of the orthonormal basis
of $\mathcal{H}_A$ and $\mathcal{H}_B$
\begin{equation}
\begin{aligned}
 |\Psi\ket=\mathcal{N}\sum_{i=1}^{d_A}\sum_{\al=1}^{d_B}X_{i\al}|i\ket_A\otimes
|\al\ket_B,
\end{aligned} 
\label{eq:psi-exp}
\end{equation}
where $\mathcal{N}$ is the normalization factor to ensure the unit norm of $|\Psi\ket$
\begin{equation}
\begin{aligned}
 \bra\Psi|\Psi\ket=1.
\end{aligned} 
\end{equation}
It is useful to regard the coefficient $X_{i\al}$ in \eqref{eq:psi-exp}
as a component of the $d_A\times d_B$ complex matrix $X$
\begin{equation}
\begin{aligned}
 X=(X_{i\al}).
\end{aligned}
\label{eq:X-mat} 
\end{equation}
Then the normalization factor $\mathcal{N}$ is written as
\begin{equation}
\begin{aligned}
 \mathcal{N}=\frac{1}{\rt{\Tr(XX^\dag)}}.
\end{aligned} 
\end{equation}

We are interested in the reduced density matrix $\rho_A$ 
defined in \eqref{eq:rhoA}
obtained by tracing out $B$. 
In terms of the matrix $X$ in \eqref{eq:X-mat}, $\rho_A$ is written as
a $d_A\times d_A$ matrix
\begin{equation}
\begin{aligned}
 \rho_A=\frac{XX^\dag}{\Tr(XX^\dag)}=\frac{W}{\Tr W},
\end{aligned} 
\end{equation} 
where  $W=XX^\dag$.  
The ensemble average over the random pure state
$|\Psi\ket$ can be defined by the Gaussian integral over the matrix
$X$
\begin{equation}
\begin{aligned}
 \bra \cO(W)\ket=\frac{\int dXdX^\dag \cO(W) e^{-\Tr(XX^\dag)}}{\int dXdX^\dag e^{-\Tr(XX^\dag)}}.
\end{aligned} 
\label{eq:O-vev}
\end{equation}
As a distribution of the matrix $W=XX^\dag$,
this is known as the Wishart-Laguerre ensemble.
See \cite{Collins2015} for a nice review on this subject.

The matrix integral \eqref{eq:O-vev} can be written as the eigenvalue integral by
diagonalizing the matrix $W$.
In the original paper by Page \cite{Page:1993df}, the entropy $S_A$ was computed by
evaluating the eigenvalue integral of $\log\rho_A$ directly.
In this paper, we will compute the entropy $S_A$ and the capacity
$C_A$ using the replica method \eqref{eq:def-SC}.
To do this, we need the expectation value of the moment $\Tr\rho_A^n$.
Fortunately, the exact result of $\bra\Tr\rho_A^n\ket$ is already 
obtained in 
\cite{Nechita2007}\footnote{
As discussed in \cite{Nechita2007}, the moments of $\rho_A=\frac{W}{\Tr W}$ 
and those of $W$ for the Wishart-Laguerre ensemble
are related by
\begin{equation}
\begin{aligned}
 \bra\Tr\rho_A^n\ket=\frac{\Ga(d_Ad_B)}{\Ga(d_Ad_B+n)}\bra\Tr W^n\ket.
\end{aligned} 
\label{eq:tr-rel}
\end{equation}
The exact result of $\bra\Tr W^n\ket$ in the Wishart-Laguerre ensemble was 
obtained
in \cite{hanlon1992some,Haagerup2003}, which is equivalent to \eqref{eq:tr-exact}
via the relation \eqref{eq:tr-rel}.
}
\begin{equation}
\begin{aligned}
\bra \Tr\rho_A^n\ket&=
\frac{\Ga(d_Ad_B)}{\Ga(d_Ad_B+n)}\cdot \frac{1}{n}\sum_{j=1}^{n}(-1)^{j-1}
\frac{[d_A+n-j]_n[d_B+n-j]_n}{(j-1)!(n-j)!}, 
\end{aligned} 
\label{eq:tr-exact}
\end{equation}
where $[a]_n=a(a-1)\cdots(a-n+1)$ denotes the falling factorial.
For instance, the first few terms of $\bra \Tr\rho_A^n\ket$ read
\begin{equation}
\begin{aligned}
 \bra\Tr\rho_A^2\ket&=\frac{d_A+d_B}{d_Ad_B+1},\\
\bra\Tr\rho_A^3\ket&=\frac{d_A^2+3d_Ad_B+d_B^2+1}{(d_Ad_B+1)(d_Ad_B+2)},\\
\bra\Tr\rho_A^4\ket&=\frac{d_A^3+6d_A^2d_B+6d_Ad_B^2+d_B^3+5d_A+5d_B}{(d_Ad_B+1)(d_Ad_B+2)(d_Ad_B+3)},
\end{aligned} 
\label{eq:full-example}
\end{equation}
which agree with the known exact results of $\bra \Tr\rho_A^n\ket$
\cite{Lubkin1978,Sommers2004}.
Note that $\bra \Tr\rho_A^n\ket$ in \eqref{eq:tr-exact} is symmetric
under the exchange of $d_A$ and $d_B$, which implies that
$S_A$ and $C_A$ are also symmetric functions of $d_A$ and $d_B$
\begin{equation}
\begin{aligned}
 S_A(d_A,d_B)=S_A(d_B,d_A),\quad
C_A(d_A,d_B)=C_A(d_B,d_A).
\end{aligned} 
\label{eq:sym-SC}
\end{equation}
In what follows, we will assume $d_A\leq d_B$ without loss of generality.
$S_A$ and $C_A$ in the opposite regime $d_A>d_B$ can be obtained
by exchanging $d_A$ and $d_B$ using the symmetry \eqref{eq:sym-SC}.

When $n$ is a positive integer,
the summation of $j$ in \eqref{eq:tr-exact} can be extended to $j=\infty$
since the summand vanishes for $j>n$.
Then \eqref{eq:tr-exact} becomes
\begin{equation}
\begin{aligned}
 \bra \Tr\rho_A^n\ket&=
\frac{\Ga(d_Ad_B)}{\Ga(d_Ad_B+n)}\cdot \frac{1}{n}\sum_{j=1}^{\infty}(-1)^{j-1}
\frac{[d_A+n-j]_n[d_B+n-j]_n}{(j-1)!(n-j)!}\\
&=\frac{\Ga(d_A+n)\Ga(d_B+n)\Ga(d_Ad_B+1)}{\Ga(n+1)\Ga(d_A+1)\Ga(d_B+1)\Ga(d_Ad_B+n)}\\
&~~~\times{}_3F_2\bigl(\{1-d_A,1-d_B,1-n\},\{1-d_A-n,1-d_B-n\};1\bigr).
\end{aligned} 
\label{eq:tr-3F2}
\end{equation}
The last expression makes sense for non-integer $n$ and it defines 
an analytic continuation of $\bra \Tr\rho_A^n\ket$ away from the integer $n$.
One can in principle compute the derivative of the
last expression in \eqref{eq:tr-3F2} with respect to $n$ to find $S_A$ and $C_A$.
However, it is not straightforward to simplify the 
derivative of the hypergeometric function ${}_3F_2$ in \eqref{eq:tr-3F2}.

It turns out that 
it is useful to rewrite \eqref{eq:tr-3F2} as the following form
using the identity of ${}_3F_2$ \footnote{
See the identity in 
the Wolfram Functions Site \href{http://functions.wolfram.com/07.27.17.0046.01}{\texttt{http://functions.wolfram.com/07.27.17.0046.01}}.}
\begin{equation}
\begin{aligned}
 \bra \Tr\rho_A^n\ket&=\frac{\Ga(d_B+n)\Ga(d_Ad_B+1)}{\Ga(d_B+1)\Ga(d_Ad_B+n)}\\
&\times{}_3F_2\bigl(\{1-d_A,1-n,-n\},\{2,1-d_B-n\};1\bigr).
\end{aligned} 
\label{eq:hyp}
\end{equation}
We find that this is expanded as
\begin{equation}
\begin{aligned}
 \bra \Tr\rho_A^n\ket=
\sum_{k=1}^{\infty}N_{n,k}\frac{\Ga(d_A)\Ga(d_B+1+n-k)\Ga(d_Ad_B+1)}{\Ga(d_A+1-k)\Ga(d_B+1)\Ga(d_Ad_B+n)},
\end{aligned} 
\label{eq:tr-narayana}
\end{equation}
where $N_{n,k}$ is the Narayana number
\begin{equation}
\begin{aligned}
 N_{n,k}=\frac{1}{n}\binom{n}{k}\binom{n}{k-1}=\frac{\Ga(n)\Ga(n+1)}{k!(k-1)!\Ga(1+n-k)\Ga(2+n-k)}.
\end{aligned} 
\label{eq:narayana}
\end{equation}
Indeed, one can show that the summation in \eqref{eq:tr-narayana}
reproduces the hypergeometric function in \eqref{eq:hyp}. 
This expression \eqref{eq:tr-narayana} makes contact with the planar limit of $\bra\Tr\rho_A^n\ket$
where the Narayana number naturally appears 
from the number of non-crossing permutations \cite{Kudler-Flam:2021rpr}.
When $n$ is a positive integer, the summation of $k$ in \eqref{eq:tr-narayana}
is truncated at $k=n$ and one can easily check that
\eqref{eq:tr-narayana} reproduces the result \eqref{eq:full-example} for small $n$.

Using the analytic continuation of the Narayana number 
by the last expression in \eqref{eq:narayana}, 
we can define a natural analytic continuation of
$\bra\Tr\rho_A^n\ket$ in \eqref{eq:tr-narayana} for non-integer $n$. 
When $d_A\leq d_B$ and $d_A$ is a positive integer, the summand 
in \eqref{eq:tr-narayana} vanishes for $k>d_A$ and hence
\eqref{eq:tr-narayana} becomes
\begin{equation}
\begin{aligned}
  \bra \Tr\rho_A^n\ket=
\sum_{k=1}^{d_A}N_{n,k}\frac{\Ga(d_A)\Ga(d_B+1+n-k)\Ga(d_Ad_B+1)}{\Ga(d_A+1-k)\Ga(d_B+1)\Ga(d_Ad_B+n)}.
\end{aligned} 
\label{eq:full-trace}
\end{equation}
In section \ref{sec:exact}, we will use this expression of $\bra \Tr\rho_A^n\ket$
for the replica computation of the exact $S_A$ and $C_A$.

\section{Planar limit}\label{sec:planar}
Before discussing the exact result of $C_A$,
in this section we will compute 
$C_A$ in the planar limit
\begin{equation}
\begin{aligned}
 d_A,d_B\to\infty~~\text{with}~~\al=\frac{d_A}{d_B}:~\text{fixed}.
\end{aligned} 
\label{eq:planar}
\end{equation}
We will assume $\al\leq1$ without loss of generality.
$C_A$ in the opposite regime $\al>1$ can be obtained by sending $\al\to\al^{-1}$
using the symmetry \eqref{eq:sym-SC}.
The computation of $C_A$ in this limit \eqref{eq:planar}
has been already done in \cite{Kawabata:2021hac}
using the planar eigenvalue density of the Wishart-Laguerre ensemble, known as the 
Marchenko–Pastur distribution.
Here we will use the replica method to compute $C_A$,
which clarifies the role of replica wormholes in $C_A$.

In the planar limit \eqref{eq:planar}, the exact result of $\bra\Tr\rho_A^n\ket$
in \eqref{eq:full-trace}
reduces to \cite{Kudler-Flam:2021rpr}
\begin{equation}
\begin{aligned}
 \bra\Tr\rho_A^n\ket_{\text{planar}}=d_A^{1-n}\sum_{k=1}^\infty N_{n,k}\al^{k-1}.
\end{aligned} 
\label{eq:tr-planar}
\end{equation}
When $n$ is a positive integer, the sum over $k$ is truncated to $1\leq k\leq n$
since the Narayana number $N_{n,k}$ in
\eqref{eq:narayana} vanishes for $k\geq n+1$.
The first few terms of the planar expectation values of $\Tr\rho_A^n$ are
given by
\begin{equation}
\begin{aligned}
 \bra\Tr\rho_A^2\ket_{\text{planar}}&=\frac{d_A+d_B}{d_Ad_B},\\
\bra\Tr\rho_A^3\ket_{\text{planar}}&=\frac{d_A^2+3d_Ad_B+d_B^2}{(d_Ad_B)^2},\\
\bra\Tr\rho_A^4\ket_{\text{planar}}&=\frac{d_A^3+6d_A^2d_B+6d_Ad_B^2+d_B^3}{(d_Ad_B)^3}.
\end{aligned} 
\label{eq:planar-example}
\end{equation}
One can see that \eqref{eq:planar-example} is obtained from the planar limit
of the exact result \eqref{eq:full-example}, as expected.

When $n$ is a positive integer, \eqref{eq:tr-planar} is expanded as
\begin{equation}
\begin{aligned}
 \bra\Tr\rho_A^n\ket_{\text{planar}}
=d_A^{1-n}+\hf n(n-1)d_A^{2-n}d_B^{-1}+\cdots +d_B^{1-n}.
\end{aligned} 
\end{equation}
In the gravitational path integral, the first term $d_A^{1-n}$ 
and the last term $d_B^{1-n}$ come from
fully disconnected and fully connected geometries, respectively.
If we assume $d_A<d_B$, the dominant contribution is the first term $d_A^{1-n}$ and 
$\bra\Tr\rho_A^n\ket_{\text{planar}}$ is written as
\begin{equation}
\begin{aligned}
 \bra\Tr\rho_A^n\ket_{\text{planar}}=d_A^{1-n}\Bigl[1+f(n,\al)\Bigr],
\end{aligned} 
\label{eq:f-sub}
\end{equation}
where $f(n,\al)$ is given by
\begin{equation}
\begin{aligned}
f(n,\al)=\sum_{k=2}^\infty N_{n,k}\al^{k-1}={}_2F_1(1-n,-n;2;\al)-1.
\end{aligned} 
\label{eq:f-def}
\end{equation}
In other words, in the gravitational picture
$f(n,\al)$ summarizes all contributions from
the sub-dominant, partially connected geometries.
Note that $f(n,\al)$ vanishes at $n=1$ by our definition in \eqref{eq:f-def}
\begin{equation}
\begin{aligned}
 f(1,\al)=0.
\end{aligned} 
\end{equation}
Plugging \eqref{eq:f-sub} into \eqref{eq:def-SC} we find
\begin{equation}
\begin{aligned}
 S_{A,\text{planar}}&=\log{d_A}-f'(1,\al),\\
 C_{A,\text{planar}}&=f''(1,\al)-f'(1,\al)^2,
\end{aligned}
\label{eq:C-f} 
\end{equation}
where the prime in $f'$ and $f''$ denotes the derivative with respect to $n$.
Note that the capacity is completely determined by the sub-leading contributions
$f(n,\al)$ in \eqref{eq:f-sub}.
In fact, if we use the leading approximation of the trace
\begin{equation}
\begin{aligned}
 \bra\Tr\rho_A^n\ket_{\text{planar}}\approx d_A^{1-n},
\end{aligned} 
\end{equation}
the capacity vanishes
\begin{equation}
\begin{aligned}
 C_{A,\text{planar}}&=
\del_n^2\bra\Tr\rho_A^n\ket_{\text{planar}}\Big|_{n=1}
-\left(\del_n\bra\Tr\rho_A^n\ket_{\text{planar}}\Big|_{n=1}\right)^2\\
&\approx(\log d_A)^2-(\log d_A)^2=0.
\end{aligned}
\label{eq:vanish-lead} 
\end{equation}
The same conclusion holds in the opposite regime $d_A>d_B$ as well
if we use the leading approximation 
$\bra\Tr\rho_A^n\ket_{\text{planar}}\approx d_B^{1-n}$.
This implies that the dominant saddle point of gravitational
path integral does not contribute to the capacity.
In other words, the capacity of entanglement is sensitive to the
sub-dominant saddle points corresponding to the partially connected
geometries. Thus, $C_A$ is a useful probe of the contributions
of replica wormholes which are not 
fully connected nor fully disconnected geometries, but some ``intermediate''
geometries.
This is consistent with the 
result in \cite{deBoer:2018mzv}
that $C_A$ takes a non-zero value for partially entangled states
and $C_A$ vanishes for the pure state or a maximally entangled state.
Namely, $C_A$ is a measure of partial entanglement \cite{deBoer:2018mzv}.

Let us evaluate $f'(1,\al)$ and $f''(1,\al)$.
From \eqref{eq:f-def} they are written as
\begin{equation}
\begin{aligned}
 f'(1,\al)&=\sum_{k=2}^\infty \del_nN_{n,k}\Big|_{n=1}\al^{k-1},\\
f''(1,\al)&=\sum_{k=2}^\infty \del_n^2N_{n,k}\Big|_{n=1}\al^{k-1}.
\end{aligned} 
\label{eq:f'-sum}
\end{equation}
Thus, we need to compute the derivative of Narayana number $N_{n,k}$ at $n=1$.
From \eqref{eq:narayana}, one can easily show that 
$N_{n,k}$ is expanded around $n=1$ as
\begin{equation}
\begin{aligned}
N_{n,k}=\left\{
\begin{aligned}
 &1, &\quad(k=1),\\
&\hf(n-1)+\hf(n-1)^2, &\quad(k=2),\\
&-\frac{(n-1)^2}{k(k-2)(k-1)^2}+\cO\big((n-1)^3\big), &\quad (k\geq3).
\end{aligned}\right.
\end{aligned} 
\label{eq:N-exp}
\end{equation}
This implies that the first derivative $\del_nN_{n,k}$ at $n=1$ vanishes unless
$k=2$, and
$f'(1,\al)$ in \eqref{eq:f'-sum} becomes
\begin{equation}
\begin{aligned}
 f'(1,\al)=\al\del_nN_{n,2}\Big|_{n=1}=
\frac{\al}{2}.
\end{aligned} 
\label{eq:f'}
\end{equation}
In a similar manner, 
from \eqref{eq:N-exp} we find that $f''(1,\al)$ in \eqref{eq:f'-sum} becomes
\begin{equation}
\begin{aligned}
 f''(1,\al)
&=\al-\sum_{k=3}^\infty \frac{2}{k(k-2)(k-1)^2}\al^{k-1}\\
&=-1-\frac{3\al}{2}+(\al-\al^{-1})\log(1-\al)+2\text{Li}_2(\al).
\end{aligned} 
\label{eq:f''}
\end{equation}
Finally, plugging the result 
of $f'(1,\al)$ in \eqref{eq:f'} and $f''(1,\al)$ in \eqref{eq:f''}
into the definition of $S_{A,\text{planar}}$
and $C_{A,\text{planar}}$ in \eqref{eq:C-f} we find
\begin{equation}
\begin{aligned}
S_{A,\text{planar}}&=\log d_A-\frac{\al}{2},\\
 C_{A,\text{planar}}&=
-1-\frac{3\al}{2}-\frac{\al^2}{4}+(\al-\al^{-1})\log(1-\al)+2\text{Li}_2(\al).
\end{aligned} 
\label{eq:C-planar}
\end{equation}
This agrees with the result in \cite{Kawabata:2021hac}
obtained from the Marchenko–Pastur distribution.
Our replica computation reveals the importance of the sub-leading contribution
$f(n,\al)$ to the capacity of entanglement.

We note in passing that $C_{A,\text{planar}}$
in \eqref{eq:C-planar} takes the maximal value at $\al=1$, or $d_A=d_B$
\cite{Kawabata:2021hac,deBoer:2018mzv}
\begin{equation}
\begin{aligned}
 C_{A,\text{planar}}^{(\text{max})}=C_{A,\text{planar}}\Big|_{\al=1}
=\frac{\pi^2}{3}-\frac{11}{4}.
\end{aligned} 
\label{eq:max-planar}
\end{equation}

\section{Exact capacity of entanglement at finite $d_A,d_B$}\label{sec:exact}

In this section we will compute the exact $S_A$ and $C_A$
using the exact result of $\bra\Tr\rho_A^n\ket$ in \eqref{eq:full-trace}.\footnote{
The exact computation of $S_A$ by the replica method 
is also considered in \cite{dyer2014divergence}
using the expression of $\bra\Tr\rho_A^n\ket$ in \eqref{eq:tr-exact}.
}
Here we assume that $d_A$ and $d_B$ are both integers and $d_A\leq d_B$.

Let us first compute the entanglement entropy $S_A$.
Plugging \eqref{eq:full-trace} into the definition 
of $S_A$ in \eqref{eq:def-SC} we find
\begin{equation}
\begin{aligned}
 S_A&=-\lim_{n\to1}\sum_{k=1}^{d_A}\frac{\Ga(d_A)\Ga(d_B+1+n-k)\Ga(d_Ad_B+1)}{\Ga(d_A+1-k)\Ga(d_B+1)\Ga(d_Ad_B+n)}\\
&\times \Biggl[N_{n,k}\Bigl(\psi(d_B+1+n-k)-\psi(d_Ad_B+n)\Bigr)
+\del_nN_{n,k}\Biggr],
\end{aligned} 
\label{eq:SA-lim}
\end{equation}
where $\psi(z)$ denotes the digamma function
\begin{equation}
\begin{aligned}
 \psi(z)=\frac{d}{dz}\log\Ga(z).
\end{aligned} 
\end{equation}
From the behavior \eqref{eq:N-exp} of the Narayana number $N_{n,k}$ near $n=1$,
\eqref{eq:SA-lim} becomes
\begin{equation}
\begin{aligned}
 S_A=\psi(d_Ad_B+1)-\psi(d_B+1)-\hf \frac{\Ga(d_A)\Ga(d_B)}{\Ga(d_A-1)\Ga(d_B+1)}.
\end{aligned} 
\label{eq:SA-psi}
\end{equation}
Using the property of the digamma function
\begin{equation}
\begin{aligned}
 \psi(m+1)=\sum_{k=1}^m \frac{1}{k}-\ga,\quad (m\in\mathbb{N}),
\end{aligned} 
\end{equation}
with $\ga$ being the Euler's constant, we arrive at the exact
entanglement entropy $S_A$ for $d_A\leq d_B$
\begin{equation}
\begin{aligned}
  S_A=
 \sum_{k=d_B+1}^{d_Ad_B}\frac{1}{k}-\frac{d_A-1}{2d_B}.
\end{aligned} 
\label{eq:SA-exact}
\end{equation}
This agrees with the famous Page's result \cite{Page:1993df}.\footnote{
The exact $S_A$ in \eqref{eq:SA-exact} is conjectured in \cite{Page:1993df}
and it is later proved in
\cite{Foong1994,Sanchez1995,Sen1996}.}
$S_A$ in the opposite regime $d_A>d_B$ is obtained from \eqref{eq:SA-exact}
by exchanging the role of $d_A$ and $d_B$.
Note that the first term of \eqref{eq:SA-exact} is written as
\begin{equation}
\begin{aligned}
 \sum_{k=d_B+1}^{d_Ad_B}\frac{1}{k}=H_{d_Ad_B}-H_{d_B},
\end{aligned} 
\end{equation}
where $H_m=\sum_{k=1}^{m}1/k$ denotes the harmonic number.

\begin{figure}[htb]
\centering
\includegraphics[width=0.7\linewidth]{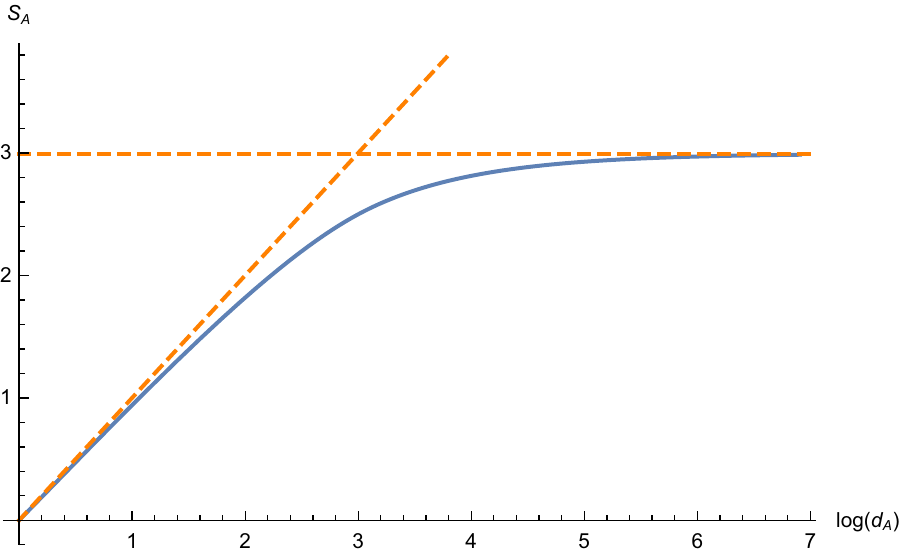}
  \caption{
Plot 
of the entanglement entropy $S_A$ as a function of $\log d_A$. We have set $d_B=20$ 
in this figure. The blue solid curve represents the exact result of $S_A$ in 
\eqref{eq:SA-exact}.
The orange dashed curve is the leading approximation $S_A=\log d_A~(d_A\leq d_B)$
and $S_A=\log d_B~(d_A>d_B)$.
}
  \label{fig:entropy}
\end{figure}

In Fig.~\ref{fig:entropy}, we show the plot of $S_A$ as a function of $\log d_A$
with a fixed $d_B$.
As we can see from Fig.~\ref{fig:entropy},
$S_A$ grows like $\log d_A$ for small $d_A$ and approaches $\log d_B$
for large $d_A$.
At least qualitatively,
this reproduces the Page curve of the Hawking radiation \cite{Page:1993wv}
if we regard the subsystem $A$ as the radiation and the subsystem $B$ as the black hole
and $t=\log d_A$ as time.
Around the Page time $t=\log d_B$, 
the contributions from the fully disconnected and the fully connected geometries exchange
dominance in the replica computation of $S_A$ 
\cite{Penington:2019kki, Almheiri:2019qdq}.

Next, let us compute the capacity of entanglement $C_A$ 
by the replica method \eqref{eq:def-SC}.
To do this, we need to compute the second derivative of $\bra\Tr\rho_A^n\ket$
at $n=1$
\begin{equation}
\begin{aligned}
 \del_n^2\bra\Tr\rho_A^n\ket\Big|_{n=1}=&
\lim_{n\to1}
\sum_{k=1}^{d_A}\frac{\Ga(d_A)\Ga(d_B+1+n-k)\Ga(d_Ad_B+1)}{\Ga(d_A+1-k)\Ga(d_B+1)\Ga(d_Ad_B+n)}\\
\times & \Biggl[N_{n,k}\Bigl(\psi(d_B+1+n-k)-\psi(d_Ad_B+n)\Bigr)^2\\
&+2\del_nN_{n,k}\Bigl(\psi(d_B+1+n-k)-\psi(d_Ad_B+n)\Bigr)\\
&+N_{n,k}\Bigl(\psi_1(d_B+1+n-k)-\psi_1(d_Ad_B+n)\Bigr)+\del_n^2N_{n,k}\Biggr],
\end{aligned} 
\end{equation}
where $\psi_1(z)=\frac{d}{dz}\psi(z)$ denotes the trigamma function.
Using the relation
\begin{equation}
\begin{aligned}
 \psi_1(m+1)=\frac{\pi^2}{6}-\sum_{k=1}^m\frac{1}{k^2},\quad
(m\in\mathbb{N}),
\end{aligned} 
\end{equation}
and the behavior \eqref{eq:N-exp} of $N_{n,k}$ near $n=1$,
after some algebra we find the exact result of capacity $C_A$
for $d_A\leq d_B$
\begin{equation}
\begin{aligned}
 C_A=\sum_{k=d_B+1}^{d_Ad_B}\frac{1}{k^2}-\frac{(d_A-1)(d_A+3)}{4d_B^2}+\frac{d_A-1}{d_B}
-\sum_{k=3}^{d_A}\frac{2}{k(k-2)(k-1)^2}\prod_{i=0}^{k-2}\frac{d_A-1-i}{d_B-i}.
\end{aligned} 
\label{eq:C-exact}
\end{equation} 
This is our main result. $C_A$ in the opposite regime
$d_A>d_B$ is obtained from \eqref{eq:C-exact} by
exchanging $d_A$ and $d_B$ using the symmetry \eqref{eq:sym-SC}.
Note that the first term of \eqref{eq:C-exact} is written as
\begin{equation}
\begin{aligned}
 \sum_{k=d_B+1}^{d_Ad_B}\frac{1}{k^2}=H_{d_Ad_B}^{(2)}-H_{d_B}^{(2)},
\end{aligned} 
\end{equation}
where $H_m^{(2)}=\sum_{k=1}^m1/k^2$ denotes the generalized 
harmonic number of the 2nd order.
This is similar to the first term of $S_A$ in \eqref{eq:SA-exact}, 
but the other terms in $C_A$ are more complicated than $S_A$.

One can easily check that \eqref{eq:C-exact} reduces to
$C_{A,\text{planar}}$ in \eqref{eq:C-planar} in the planar limit \eqref{eq:planar}.
Also, one can check that $C_A(d_A,d_B)$ in \eqref{eq:C-exact}
for $d_A,d_B=2,3$ agree with the result in \cite{deBoer:2018mzv}
\begin{equation}
\begin{aligned}
 C_A(2,2)&=\frac{13}{36},\\
C_A(2,3)&=\frac{1169}{3600},\\
C_A(3,3)&=\frac{2898541}{6350400}.
\end{aligned} 
\end{equation}
From the exact result \eqref{eq:C-exact},
we find the small $d_A$ and the large $d_A$ behavior of $C_A$
\begin{equation}
\begin{aligned}
 C_A\approx \left\{
\begin{aligned}
&\left(d_A-\frac{1}{d_A}\right)\frac{1}{d_B},&\quad &(1\leq d_A\ll d_B),\\
 &\left(d_B-\frac{1}{d_B}\right)\frac{1}{d_A},&\quad &(d_A\gg d_B).
\end{aligned}
\right.
\end{aligned} 
\label{eq:CA-asym}
\end{equation}
Our exact $C_A$ in \eqref{eq:C-exact}
takes the maximal value at $d_A=d_B$
\begin{equation}
\begin{aligned}
 C_A^{(\text{max})}=\sum_{k=1}^{d_B^2}\frac{1}{k^2}+
\sum_{k=1}^{d_B}\frac{1}{k^2}+\frac{1}{d_B}-\frac{1}{4d_B^2}-
\frac{11}{4}.
\end{aligned} 
\end{equation}
In the large $d_B$ limit this is expanded as
\begin{equation}
\begin{aligned}
 C_A^{(\text{max})}=\frac{\pi^2}{3}-\frac{11}{4}-\frac{3}{4d_B^2}+\cO(d_B^{-3}),
\end{aligned} 
\end{equation}
where the first two terms agree with the maximal value of capacity in 
the planar limit \eqref{eq:max-planar}.
For finite $d_A,d_B$, 
one can show that the exact $C_A$ is bounded from above
\begin{equation}
\begin{aligned}
 C_A\leq C_A^{(\text{max})}<
\frac{\pi^2}{3}-\frac{11}{4}.
\end{aligned} 
\end{equation}

\begin{figure}[htb]
\centering
\includegraphics[width=0.7\linewidth]{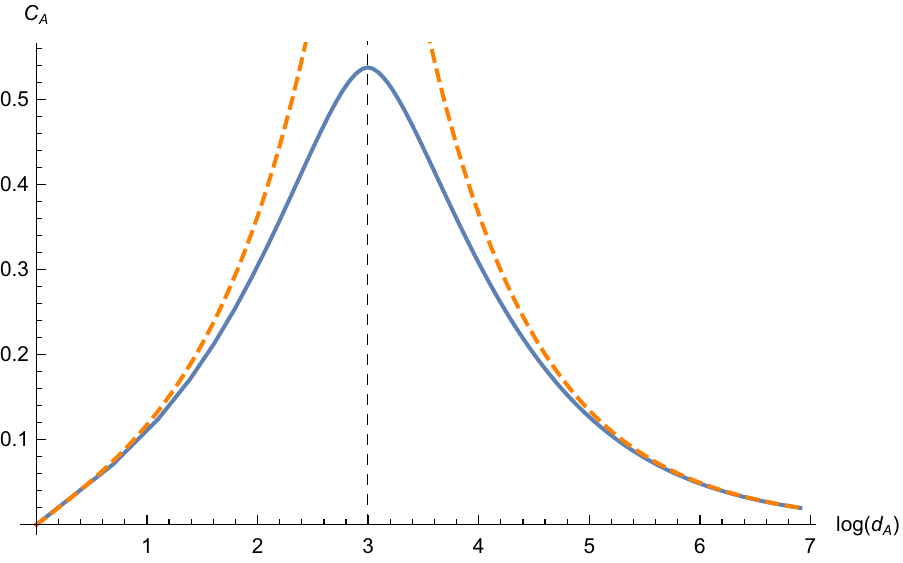}
  \caption{
Plot 
of the capacity of entanglement $C_A$ as a function of $\log d_A$. We have set $d_B=20$ 
in this figure.
The blue solid curve is the exact result of $C_A$ in \eqref{eq:C-exact}.
The orange dashed curves represent 
the asymptotic behavior of $C_A$ in \eqref{eq:CA-asym}.
The dashed vertical line is at $d_A=d_B$ where $C_A$ becomes maximal.
}
  \label{fig:capacity}
\end{figure}
In Fig.~\ref{fig:capacity}, we show the plot of the exact capacity
$C_A$ in \eqref{eq:C-exact}.
We can see that $C_A$ vanishes for $d_A=1$ and 
approaches zero at large $d_A\gg d_B$.
This is qualitatively similar to the result of the planar limit found in
\cite{Kawabata:2021hac}.
We emphasize that our result \eqref{eq:C-exact}
is exact at finite $d_A,d_B$ and \eqref{eq:C-exact}
includes all the non-planar corrections.
As we argued in the previous section, $C_A$ is sensitive
to the sub-leading terms in $\bra\Tr\rho_A^n\ket$
corresponding to the partially connected geometries.
Indeed, $C_A$ vanishes at the early and late ``time'' $t=\log d_A$ where
the fully connected or fully disconnected geometry is dominant.
$C_A$ takes a non-zero value near the Page time $t=\log d_B$ (or $d_A=d_B$)
which is interpreted that 
the partially connected geometries give substantial contributions to $C_A$
near the Page time.

\section{Conclusions and outlook}\label{sec:conclusion}
In this paper, we have computed 
the exact capacity of entanglement $C_A$ \eqref{eq:C-exact}
at finite $d_A,d_B$ using the replica method \eqref{eq:def-SC}.
At the technical level, the important ingredient
in our computation is the new exact formula \eqref{eq:full-trace} 
of $\bra\Tr\rho_A^n\ket$
written in terms of the Narayana number $N_{n,k}$.
This formula \eqref{eq:full-trace} makes the relation to the planar limit manifest.
We argued that $C_A$ vanishes for the fully connected or fully disconnected
geometries, and $C_A$ is sensitive to the sub-leading contributions
to $\bra\Tr\rho_A^n\ket$ coming from the partial connected geometries
in the gravitational path integral.
This suggests that $C_A$ is a good probe of the partial entanglement, 
as discussed in \cite{deBoer:2018mzv}.

There are several open questions. 
The capacity of entanglement is introduced
in \cite{qi} as an analogue of the heat capacity.
Indeed, if we introduce the modular Hamiltonian
$K=-\log\rho_A$, the moment $\Tr\rho_A^n$ looks like the partition function
\begin{equation}
\begin{aligned}
 Z_n=\Tr\rho_A^n=\Tr e^{-nK},
\end{aligned} 
\end{equation}
and $n$ plays the role of the inverse temperature $\bt$.
In this picture,
our definition \eqref{eq:def-SC}
of $S_A$ and $C_A$ is based on the ``annealed'' free energy $\log\bra Z_n\ket$
\begin{equation}
\begin{aligned}
 S_A=-\del_n\log \bra Z_n\ket\Big|_{n=1},\qquad
C_A=\del_n^2\log \bra Z_n\ket\Big|_{n=1}.
\end{aligned} 
\end{equation}
One could consider
the quenched version of $S_A$ and $C_A$ as well
\begin{equation}
\begin{aligned}
 S_A^{\text{qu}}=-\del_n\bra\log Z_n\ket\Big|_{n=1},\qquad
C_A^{\text{qu}}=\del_n^2\bra\log Z_n\ket\Big|_{n=1}.
\end{aligned} 
\end{equation}
We leave the computation of the quenched version of $S_A$ and $C_A$ as an interesting 
future problem.

It would be interesting to study the gravitational picture of the
capacity of entanglement. 
$C_A$ is related to the quantum fluctuation of the modular Hamiltonian and
the prescription of the gravitational computation of
$C_A$ is proposed in \cite{Nakaguchi:2016zqi}. 
We have argued that $C_A$ receives contributions only 
from the sub-leading partially connected geometries
in the replica computation. It would be interesting to related this
picture to the prescription in \cite{Nakaguchi:2016zqi}.

\acknowledgments
This work was supported in part by JSPS KAKENHI Grant No. 19K03845.
\bibliography{cite}
\bibliographystyle{utphys}

\end{document}